\documentclass[journal]{IEEEtran}
  \usepackage{graphicx}
\usepackage{amsmath}
\usepackage{algorithmic}

\usepackage[caption=false,font=footnotesize]{subfig}
\begin{document}
%
\title{Wide Area Measurement System-based Low Frequency Oscillation Damping Control through Reinforcement Learning}
%
%
%

\author{Yousaf~Hashmy,~\IEEEmembership{Student Member,~IEEE,}
        Zhe~Yu,~\IEEEmembership{Member,~IEEE,}
        Di~Shi,~\IEEEmembership{Senior Member,~IEEE,}
        and~Yang~Weng,~\IEEEmembership{Member,~IEEE}\vspace{-6mm}
        
\thanks{This work was funded by SGCC Science and Technology Program under project ``AI based oscillation detection and control” and contract number SGJS0000DKJS1801231.}
\thanks{Y. Hashmy is with GEIRI North America, San Jose,
CA,  95134, USA and Department of Electrical Computer and Energy Engineering, Arizona State University, Tempe, AZ, 85281 USA E-mail: shashmy@asu.edu.}
\thanks{Z. Yu and D. Shi are with GEIRI North America, San Jose,
CA,  95134, USA E-mails: zhe.yu@geirina.net; di.shi@geirina.net.}
\thanks{Y. Weng is with Department of Electrical Computer and Energy Engineering, Arizona State University, Tempe, AZ, 85281 USA E-mail: yang.weng@asu.edu.}
}

\maketitle


\begin{abstract}
Ensuring the stability of power systems is gaining more attraction today than ever before, due to the rapid growth of uncertainties in load and renewable energy penetration. Lately, wide area measurement system-based centralized controlling techniques started providing a more flexible and robust control to keep the system stable. But, such a modernization of control philosophy faces pressing challenges due to the irregularities in delays of long-distance communication channels and the response of equipment to control actions. Therefore, we propose an innovative approach that can revolutionize the control strategy for damping down low-frequency oscillations in transmission systems. The proposed method is enriched with the potential of overcoming the challenges of communication delays and other non-linearities in wide area damping control by leveraging the capability of the reinforcement learning technique. Such a technique has a unique characteristic to learn on diverse scenarios and operating conditions by exploring the environment and devising an optimal control action policy by implementing policy gradient method. Our detailed analysis and systematically designed numerical validation prove the feasibility, scalability and interpretability of the carefully modelled low-frequency oscillation damping controller so that stability is ensured even with the uncertainties of load and generation are on the rise.
\end{abstract}


%
\IEEEpeerreviewmaketitle

\vspace{-4mm}
\section{Introduction}

With the rapid growth of uncertainties in the power generation as well as in the load, the transmission system control faces serious concerns. Such concerns can be detrimental to the smooth operation and may result in blackouts, such as there are a series of cases of blackouts because of inter-area oscillations reported at tie-line between Pakistan and Iran \cite{zahid2017inter}. The inter-area oscillations often face poor damping, because of power systems generally operate close to their maximum available transfer capability \cite{azad2013damping} and \cite{klein1991fundamental}. 

Therefore, power engineers are working at damping down such oscillations by traditional localized power system stabilizers. But, inter-area modes are neither always controllable nor observable from local measurement signals \cite{aboul1996damping}. Therefore, with the development of wide area measurement systems and practical implementation of phasor measurement units, there has been the natural progression to use wide area damping controllers through long-distance signal transmission \cite{8386650,zhang2013design} and \cite{910791}. Hence, engineers shift to the more flexible control methodology of wide area-based damping control \cite{younis2013wide}, \cite{6161353} and \cite{yao2014wide}, so that observability of the system can be enhanced.

Moreover, there have been attempts to develop Wide Area Damping Control systems through different communication channels \cite{younis2013wide}. But there exist challenges of variable time delays and model uncertainties. All such challenges greatly impede the performance of the control mechanism. Linear control methods have been conventionally used in the past, like \cite{5275875} uses Lyapunov stability theory and linear matrix inequality (LMI) method to analyze the impact of time delays on power system stability. Moreover, \cite{gu2003stability} and \cite{6475217} follows a similar approach by extending to the Lyapunov-Krasovskii functional method to calculate the factor for compensating timing delays in the communication. Similarly, \cite{wu2004evaluation} considers the timing delay uncertainties in low-frequency oscillation damping control signals as well, but it assumes the delays as part of system uncertainties and embeds the compensation in the controller. Whereas, the performance of methods in \cite{6475217}, \cite{7053543} and \cite{wu2004evaluation} are highly sensitive to the system operating conditions, that greatly threatens the reliability of such a scheme. A two-level hierarchical controller based on wide area measurements is presented in  \cite{okou2005power}. Here the local controllers are controlling the damping of oscillations and a centralized controller oversees the performance of the local controller. However, its drawback is its reliance on building up two different controllers with more complex hardware and added cost. The large-scale power system is not only a complex, non-linear and high-order dynamic system, but it is also difficult to obtain its full scale detailed model. However, for the design of a damping controller, a low-order equivalent linear model, whose structure is fixed and parameters vary with the operating conditions, is estimated by a recursive least squares algorithm with a changing forgetting factor in  \cite{4334961}.

Moreover, \cite{ni2002power} proposes a control strategy that relies on distributed artificial intelligence. In such a method, Supervised Power System Stabilizer utilizes fuzzy logic for building up a control system that can accommodate the communication channel irregularities. But it fails to explicitly take into account, the non-linearities of the loads and generation patterns, especially when the uncertainties are inducted into the system by Distributed Energy Resources (DERs) \cite{ZHANG201645} and \cite{4113518}.

To have a better model for the stochastic nature of uncertainties stochastic control method is adopted by \cite{zhang2015stability}, but the major drawback is that it assumes long stochastic time delay, which is an unfair assumption for the power systems. To overcome that, \cite{7008569} uses the expectation modelling of time delays for shorter stochastic time delays. Additionally, \cite{yao2014wide} employs a network predictive control scheme for estimating the time delays in the communication channels which is nothing more than an extension of \cite{wu2004evaluation}, as it assumes a low-order model for a more complex power system through least-squares based identification algorithm \cite{4334961}. The compensation of time delays is carried out through a transfer function ignoring the system behavior and its dynamics, mostly. Predominantly, the timing delays are dependent on the multiple factors including the system as well as those which are not dependent on the system, so their analytical solution is hard to determine, with high confidence. The stochasticity is key while damping down the oscillations in the power system. 

To overcome the challenges in the previous methods, we propose to model the controlling agent that can provide a stochastic control through wide area measurements, so that we can ensure the extended observability and flexibility to achieve centralized control. Moreover, we model the stochastic nature of communication timing delays by introducing the randomness through the exploration of the system operating conditions. Such a method is enriched with a reinforced learning mechanism that can effectively learn the distributions of uncertainties caused by the variability of load and DERs. 

Reinforcement learning is introduced by designing an appropriate reward function, with a capability to assign a high reward for the learning agent when the low-frequency oscillations (LFOs) are damped down effectively. LFO damping requires a continuous action space; otherwise, the discretized actions will not be able to produce a smooth output. Thus, to achieve the best action set, named as policy, we use the gradient directly on the policy; otherwise, the action-value function (Q value) needs the discretized action output. However, there can be infinite possible state actions pairs, when a continuous control is required for damping down the oscillations. To solve such an issue, we use the deep neural networks with a finite number of parameters, to be learnt.

\begin{figure}[t] 
\includegraphics[width=\linewidth]{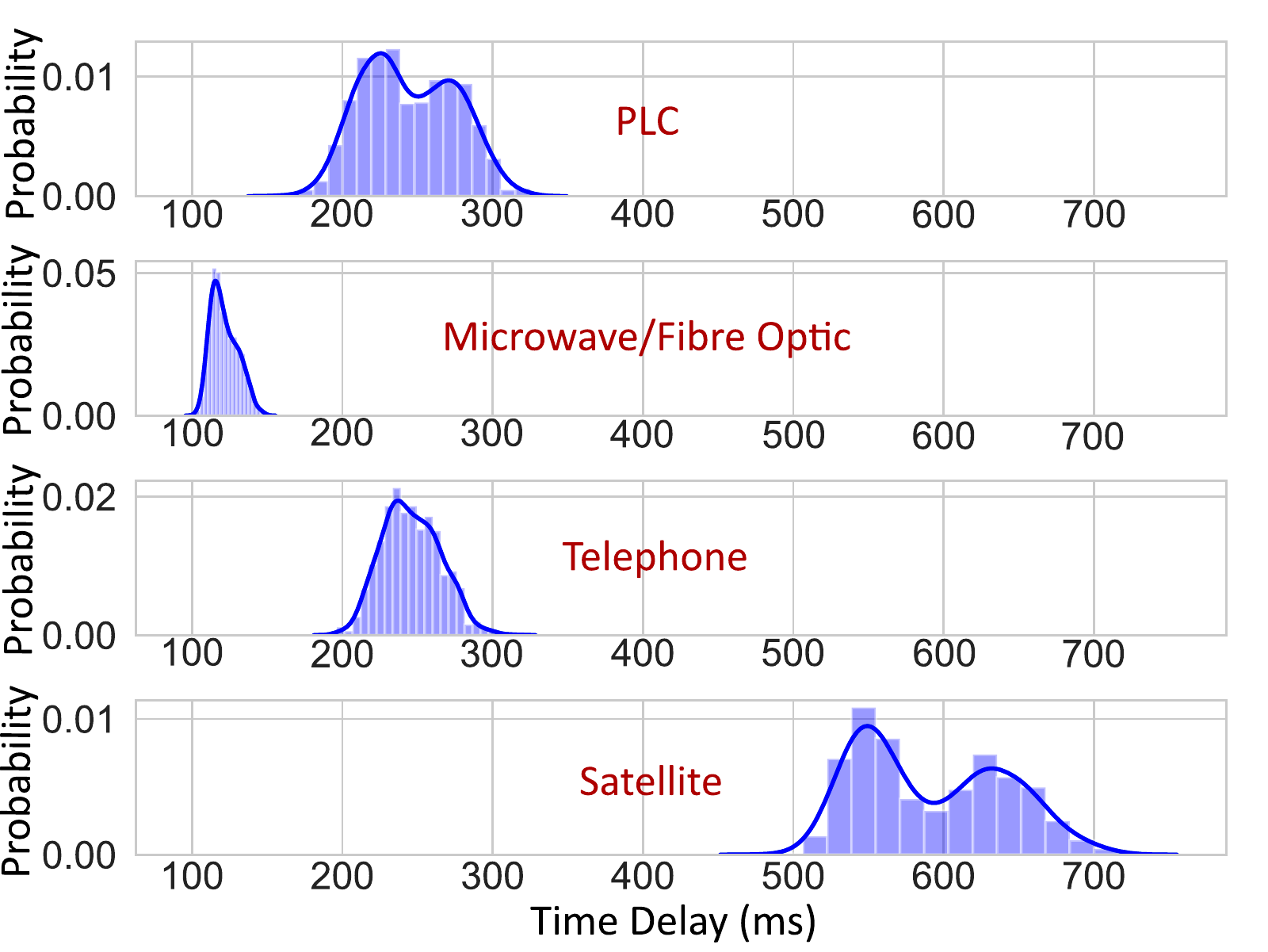}
 
  \caption{ Probability distributions of different communication links are assumed to be having a mixed Gaussian distribution.}

  \label{fig_rew_act} 
\end{figure}

The performance is evaluated on the $6$-bus system and $39$-bus New England system to prove the concept, that it is not only valid for the smaller test cases but is also applicable on larger systems in the North East of the United States of America. We assume the timing delays to follow Gaussian mixture distributions illustrated in Fig. \ref{fig_rew_act} in the range suggested by past research as in Table \ref{Table2}. Such a systematic validation shows the applicability and scalability of the reinforcement learning-based oscillation damping controller.

This paper is organized as follows: Section II is about the past control methods, section III  gives modelling and defining problem statement, Section III is about the improved control scheme with specialized reward design and the learning agent design and Section IV gives validation of the proposed methodology under different scenarios. Finally, Section V concludes the paper.
\begin{table}[!hbt]

\caption{Delays due to different communication links \cite{younis2013wide}.}

    \label{tab:timedelays}
    \centering
 \begin{tabular}{c  c  c}   

 \hline\hline
Communication Link & One way delay (ms) \\ [0.5ex] 
 \hline
 Fibre-optic cables & $\approx$ 100-150 \\ 
\hline
 Microwave links & $\approx$ 100-150 \\
\hline
 Power line carrier (PLC) & $\approx$ 150-350 \\
 \hline
 Telephone lines & $\approx$ 200-300  \\
 \hline
Satellite link & $\approx$ 500-700 \\  
 \hline\hline
\end{tabular}
\label{Table2}
\end{table}

\section{Control Method Selection}
The control theory in the recent past has shown significant developments, providing multiple control mechanisms for multi-faceted and complex environments like transmission systems. But not all of them are suited to resolving the challenge of controlling a wide area system to damp down the LFOs, in the presence of uncertainties in the system, through flexible and robust mechanism. So, we explore numerous possible candidates that can provide robust control using wide area measurements.

One of the past approaches is deterministic control, that has been applied to the power systems in abundance. But such control has a limitation of excessive reliance on the system parameters and operating conditions. Both of them are not easily observable for the power system which is extending over continents and sometimes beyond them. Moreover, the predictability of such conditions is a hard problem due to accessibility constraints, rapid growth and uncertainties created by DER integration. Hence, controlling a wide area system demands a mechanism with a potential of incorporating the uncertainties of the system.

To incorporate the system uncertainties and variabilities, a logical progression would be towards the branch of stochastic control systems. Such a control mechanism is capable of dealing with uncertainties in the observations as well as the ones due to noise. There are two distinct methods of predictive controls stochastic control namely, robust model predictive control, which is a more conservative method considering the worst scenario in the optimization procedure. However, that technique deteriorates the overall controller's performance and also is applicable only for systems with bounded uncertainties. On the contrary, a transmission system is in a constant mode of upgradation with modern equipment and renewable energy penetration into the grid, and it is virtually impossible to record all such transformations beforehand to evaluate their impact on the bounds of uncertainties. 

The alternative method is commonly known as stochastic model predictive control. Such a method considers soft constraints which limit the risk of violation by a probabilistic inequality. But such a control scheme is limited to consider the real-time events of the power system, and its inability to broaden its scope to consider the events in the past for devising a control action will never enable it to improve as well as adapt to the dynamic nature of the transmission network. 

Consequently, the past methods are not capable of providing robust control solution to the environment such as that of a transmission grid. Therefore, we propose a scheme that helps to enrich the information available by providing enough exploration of the system operation and use such historical data to provide damping control in wide area systems, without losing the stochasticity for learning uncertainties. To achieve this goal, we combine randomness with decision maker's past experiences. Such an approach is well summarized as a Markov Decision Process (MDP), with each state of the wide area system as a Markov Decision State (MDS). Such a definition enables us to introduce a stochastic variant of optimal control problem and gain the advantage of the system dynamics by enhancing our information based on measurement signals and generated action.
\section{Modelling}
Thanks to the implementation of WAMS, the observability of the system is greatly improved. Here, we assume that each generator is equipped with a Phasor Measurement Unit (PMU) so that its phase angle and speed are available to the  controller, with possible delays. Define the state $s_{t}$ for all observable generators $g=1,\cdots,G$ to be controlled. The deviations in generator speeds are $\omega_{g}^{t}$ and the phase angles are $\theta_{t}^{b}$ between the voltages of the buses $b=1,\cdots,B$ at remote locations for time $t = 1,\cdots,T$. As the speeds of generators vary upon the occurrence of disturbance, we use those deviations of the speeds $\Delta\omega^{t}_{g}=|\omega^{t}_{g}-\omega^{t-1}_{g}|$ to define state. 
\begin{equation}
\begin{aligned}
    s_{1}^{t} &= \{\Delta\omega_{1}^{t},\Delta\omega_{2}^{t},\Delta\omega_{3}^{t},\cdots,\Delta\omega_{G}^{t}\},\\
 s_{2}^{t} &= \{\theta_{1}^{t},\theta_{2}^{t},\theta_{3}^{t},\cdots,    \theta_{B}^{t}\},\\
 s_{t} &= s_{1}^{t} \cup s_{2}^{t}.
\end{aligned}
\end{equation}

The modern-day PSS is responsible for damping down LFOs by adjusting the voltage applied at the field windings $V_{g}$ of all the synchronous generators $g$. So, the output of the controller will essentially be an action vector $a_{t}$ for all the generators $g$ at time $t$. The action vector $a_t$, defined in equation (2), is a stabilizing voltage parameter that alters field voltage of synchronous generators.
\begin{equation}
\hspace{-7mm}
    a_{t} = \{V_{1}^{t},V_{2}^{t},V_{3}^{t},\cdots,V_{G}^{t}\}.
\end{equation}
States and actions enable us to completely define the problem.
\vspace{-4mm}
\subsection{Problem Definition}
\textbf{Problem}: 
damp down low-frequency inter-area oscillations by adjusting field voltages of generators.

\textbf{Given}:
\begin{itemize}
\item a transmission system as the environment $X$, 
\item state of the system $s_{t}$ in $X$, and communication time delay probability distribution $P(t_{d})$. 
\end{itemize}

\textbf{To Find}: 
\begin{itemize}
\item discounted reward $R(s_{t},a_{t})$, 
\item and a policy $\pi$ comprising of action set $a_{t}$ for stabilizing field voltages $V_{g}^{t}$ of synchronous generators.
\end{itemize}

\begin{figure}[t]
\includegraphics[width=3.7in]{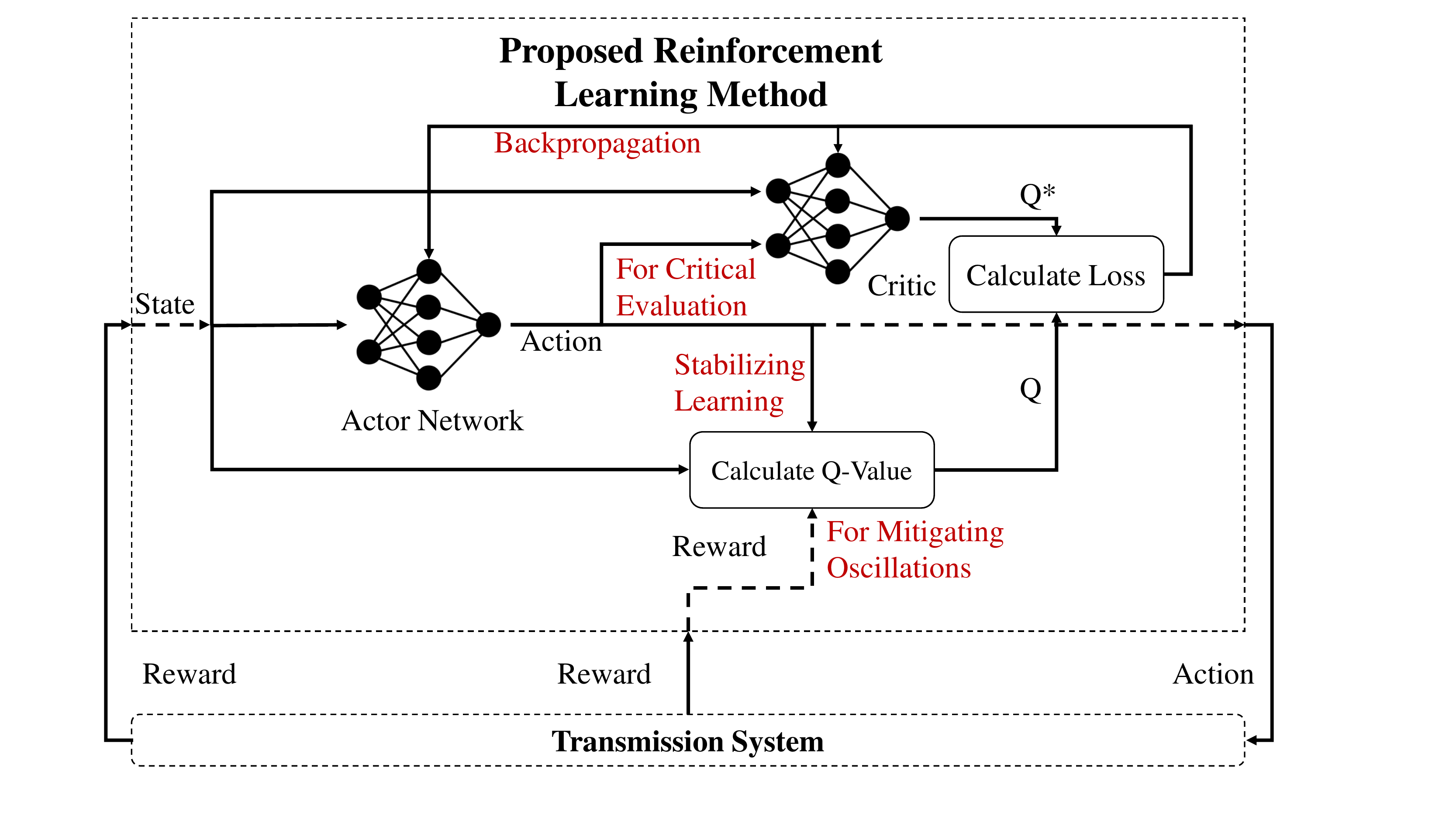}
\centering
\vspace{-4mm}
\caption{The architecture of Deep Deterministic Policy Gradient method.}
\vspace{-2mm}
\label{fig:ddpg100}
\end{figure}

\section{Control Scheme}
With a model, we aim to develop a reinforcement learning-based robust control scheme. But it is not possible unless we have a specialized reward design that can maximize the potential of the power system knowledge. 

\subsection{Reward Design with Maximized Information Gain}
With states, actions and policy clearly defined, we require an evaluation function, that helps in deciding the extent of the fidelity of generated control action. The evaluation function is also called as a reward function in the reinforcement learning domain. We design reward that can help maximize the information obtained from the wide area based observations from synchrophasors and local measurements from the generators. Our goal is to minimize the oscillations in the frequencies of wide area systems. So we propose to capture all the features related to oscillations like deviation of generator speed from $1$ pu and the abrupt changes in the generator speed with respect to time. But, that will not be enough to capture the effect of buses connected by long-distance lines, so we improve the information gain by leveraging the wide are measurements of phase angle variation between remote buses as well. We incorporate all that information into the reward design. 

Moreover, the learning agent may require an immensely large action space, impeding its performance and speed. Therefore, we further boost the information gain by incorporating the knowledge from locally used Power System Stabilizers (PSS) for each generator $g$. Such information is embedded into the reward in the form of the bounds $u$ and $v$, where $u$ is indicative of the upper bound of the control action space. Similarly, $v$ represents its lower bound. Such a definition enables the model to treat the action inside and outside the constraints separately.

The speeds of generators are perturbed upon creating the fault in any system and the low-frequency oscillations reside inside the signal are supposed to be damped down. For that reason, the reward function consists of four terms. The first terms help to bring the speed $\omega_{g}^{t}$ as close to $1$ pu as possible. Second terms overcome the sustaining deviations in the speeds of the generators. These terms are leveraging the information of the conventional local PSSs. The third terms in equations (\ref{eqn7}) and (\ref{eqn8}) refers to the operation of localized PSS which helps in limiting the bounds of actions. Whereas, the fourth terms incorporate the difference between the phase angles of voltages at remote buses. The control actions are chosen based on the oscillations between the remote buses. We use the difference of the phase angles of remote buses to increase the observability because angle differences of remote buses were unobservable without wide area damping controller. We intend to reduce such a difference so that deviations of speeds of generators connected to remote buses are limited. For all the terms we take absolute values to capture only the absolute difference so that the highest attainable reward is $0$.

As we have established the characteristics that capture the maximum information, we aim to combine them. A linear relationship among them is the most suitable one since we need to have the differences large enough where the learning agent can improve by gaining a reasonably large reward. If the reward values are too small like having higher-order terms then the learning agent will keep oscillating instead of gaining a substantially large reward. So we summarize the discussion mathematically as, 
\begin{equation}
   r(s_{t},a_{t})=%
   \begin{cases}
     r_{1}(s_{t},a_{t}) &\text{if} ~~a_{t}<u, \\
     r_{2}(s_{t},a_{t}) &\text{if}~~ a_{t}>v,\\
     r_{3}(s_{t},a_{t}) &\text{Otherwise}.
   \end{cases}
   \label{rew1}
\end{equation}

\begin{equation}
\begin{aligned}
r_{1}(s_{t},a_{t})&=\sum_{b=1}^{B}\sum_{t=0}^{T}\sum_{g=1}^{G}(-\alpha|1-\omega_{g}^{t}|-\beta|\Delta\omega_{g}^{t}|\\
&~~~-\eta|-a_{g}^{t}-u|-\zeta|\theta_{b}^{t}-\theta_{b+1}^{t}|), 
\end{aligned}
\label{eqn7}
 \end{equation}
 \begin{equation}
\begin{aligned}
r_{2}(s_{t},a_{t})&=\sum_{b=1}^{B}\sum_{t=0}^{T}\sum_{g=1}^{G}(-\alpha|1-\omega_{g}^{t}|-\beta|\Delta\omega_{g}^{t}|\\
&~~~-\eta|a_{g}^{t}-v|-\zeta|\theta_{b}^{t}-\theta_{b+1}^{t}|),
\end{aligned}
\label{eqn8}
 \end{equation}
 
\begin{equation}
\begin{aligned}
\hspace{1mm}
r_{3}(s_{t},a_{t})&=\sum_{b=1}^{B}\sum_{t=0}^{T}\sum_{g=1}^{G}(-                                     \alpha|1-\omega_{g}^{t}|-\beta|\Delta\omega_{g}^{t}|\\
&~~~-\zeta|\theta_{b}^{t}-\theta_{b+1}^{t}|),
\end{aligned}
\label{eqn9}
\end{equation}
where $\alpha$, $\beta$, $\eta$ and $\zeta$ are the scaling factors that depend greatly on the system. All of these parameters can be tuned by cross-validating the performance of the agent.

We aim to use future reward values to determine the state and action pairs resulting in a maximum expected reward. So we require the rewards of the future and sum them all, but this creates a mathematical difficulty of getting an infinite reward. This condition may happen in the power system control domain very frequently because there exists an infinite number of possible states when we aim to continuously control the generator field voltages. We avoid such a situation by using discounted reward, which gives lower weights to the rewards associated with the state and action pairs in further future. Such a definition enables us to give higher weights to the reward values which are in the near future. Hence, we can represent the discounted future reward as $R_t$ as
\begin{equation}
\begin{aligned}
R_t&= r(s_{t},a_{t})+\gamma r(s_{t+1},a_{t+1})+\gamma^{2}r(s_{t+2},a_{t+2})+\cdots,
\end{aligned}
\label{reward}
\end{equation}
that depends on a discount factor $\gamma \in [0, 1]$. The main objective of reinforcement learning is to maximize the expected discounted future reward. Therefore, we need to have a model with such functionality, under time and computational limitations.

\subsection{Learning Agent Achieving Accurate Control}
As state $s_{t}$ is involved in defining state value function $V(s_{t})$. But we intend to incorporate more information, hence we propose to use $Q(s_{t},a_{t})$ so that not only states but also the action values are considered while making the optimal control decision.
For a learning agent, we propose to use $Q(s_{t},a_{t})$ so that not only states but also the action values also considered while making the optimal control decision,
\begin{equation}
Q(s_{t},a_{t})=E[{R_{t}|s_{t},a_{t}}],
\label{q}
\end{equation}
where $R_{t}$ is the discounted reward, which is already defined in equation \ref{reward}. 

The proposed learning agent is expected to learn the optimal policy $\pi$ taking the communication delay into account. It is important because there can be uncertainties in the response of generating units and communication channel delays. Such changes become worse with the aging of the equipment. Therefore, we use the learning agent that can learn through interacting with the power system and eventually learns to deliver the policy based on maximized information. Unlike a conventional feedback control system designed to respond in a certain fashion based on the current state of the system, we propose to embed the knowledge from the past and random exploration of the transmission network so that the uncertainties due to load switching, capacitor bank switching and DERs can be learnt effectively. Moreover, the adaptation of the learning agent to the ever-dynamic nature of the power system is also significant because static parameters of the controller will become useless in a very small amount of time.

To accomplish the goals, the discretized states and actions will help to generate discrete $Q$ function, but in a real scenario such as a power system, the number of states and action pairs can grow rapidly with increasing the count of the buses in the system. This may create a $Q$ table of an infinite number of entries. To overcome this challenge, we introduce a function approximation method based on neural networks to provide a finite set of parameters that can be learnt through experience data. The data comprises of the states and actions of the current and future states along with their respective reward values. Then, we are in a better position to approximate the $Q$ function.

However, using only a single neural network might fail because there will be a high chance of falling victim to local minima. Therefore, as shown in Fig. \ref{fig:ddpg100} we choose an actor-critic model, where a critic-network might help to suppress the bad decisions made by an actor-network. The details of continuous control through the actor-critic model is presented in \cite{lillicrap2015continuous}. Our proposed model not only deals with continuous action space, but its fidelity for learning under complex environments is proven in \cite{mnih2015human}. By combining expressions (\ref{q}) and (\ref{reward}), we obtain
\begin{equation}
    Q(s_{t},a_{t}) ={E}[r(s_{t},a_{t})+\gamma \max_{a'}E[Q(s_{t+1},a_{t+1})].
\end{equation}

We evaluate the function approximator $Q$ for critic network, by sampling states from the wide area measurements following a specific distribution. An actor-network $\mu(s_{t+1})$ takes only the states as input features and directly estimates the actions. But such estimation requires critical evaluation. So, we define the approximator $Q^{*}(s_{t},a_{t})$ for critic network,
\vspace{-2mm}
\begin{equation}
\vspace{-2mm}
    Q^{*}(s_{t},a_{t}) =E[r(s_{t},a_{t})+\gamma Q^{*}(s_{t+1},\mu(s_{t+1})].
\end{equation}

Since both of the function approximators are characterized as deep layers of neural networks, we parameterize them with $\theta^{Q}$ and $\theta^{\mu}$. So, the loss function for $M$ samples becomes
\vspace{-2mm}
\begin{equation}
\vspace{-2mm}
   \textrm{Loss}=\frac{1}{M}\sum_{i=1}^{M}(Q^{*}(s_{i+1},a_{i+1})-Q(s_{i},a_{i}))^{2}.
\end{equation}
where $i=1\cdots,M$ is the number of samples in mini-batch. We obtain the parameters of critic-network by iteratively minimizing the above loss function. Whereas, we update the actor parameters using the policy gradient approach,
\vspace{-2mm}
\begin{equation}
\vspace{-2mm}
   \nabla_{\theta_{\mu}}J\approx\frac{1}{M}\nabla_{a} Q(s_{i},a_{i}|\theta^{Q}) \nabla_{\theta_{\mu}}\mu(s_{i}|\theta^{\mu}),
\end{equation}
Then we update the target actor $Q'$ and target critic $\mu'$ parameters using a periodic approach, so that after each iteration the target actor becomes the initial actor and target critic becomes the initial critic, $
\theta^{Q'}=\theta^{Q}$ and $
\theta^{\mu'}=\theta^{\mu} 
$ respectively.

\begin{figure}[t]
\includegraphics[width=3.5in]{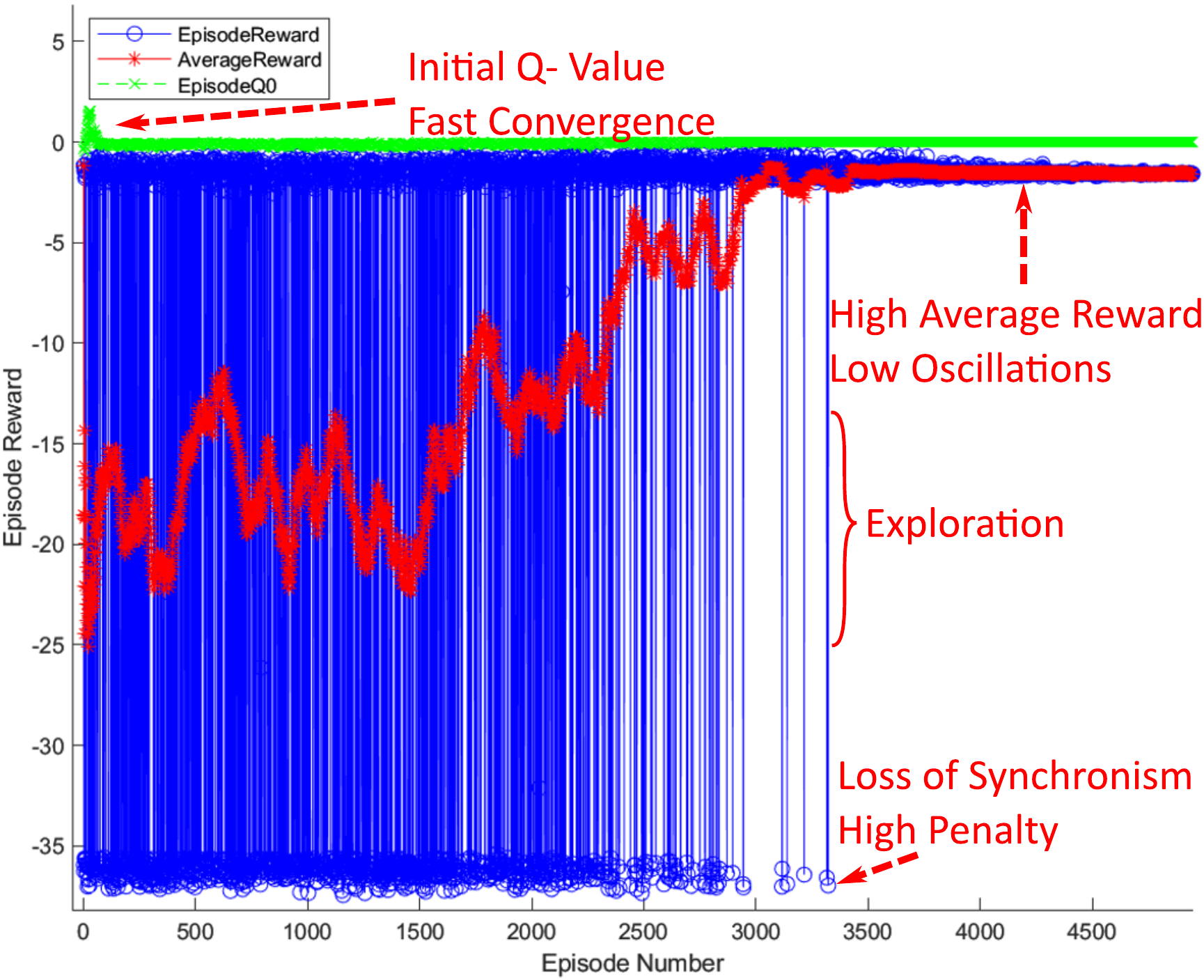}
\centering
\vspace{-6mm}
\caption{Learning curve showing reward with respect to increasing episodes. The blue points show the episode reward. Red line is showing the average reward considering an averaging window of past five episodes. Green points indicate the initial $Q$ value for each episode.}
\vspace{-4mm}
\label{fig:ddpg10}
\end{figure}

The flow of the algorithm for damping the low-frequency oscillations using this procedure can also be observed in Fig. \ref{fig:ddpg10}. The hyperparameters include discounting factor, exploration epsilon, experience buffer size and minibatch size, that require extensive cross-validation of the model on the real testing environment to reach to their optimal values. The learning model has to be accurate, but its training time is also important. Therefore, we provide a study based on the speed measurement for learning using the mean overall speed of the agent towards the target, which is defined using the success rate. The success rate is the rate at which the episodes end without the loss of synchronism of the system. 
\vspace{-2mm}
\begin{equation}
\vspace{-2mm}
\bar{v}=\frac{\sum v_{x}cos(\theta_{\textrm{target}})}{v_{x}^{\textrm{max}}},
\end{equation}
where $\bar{v}$ is the average speed, $v_{x}$ is the speed value for all the time steps in an episode. The maximum value is named as $v_{x}^{max}$. $\theta_{\textrm{target}}$ gives the parameters of the target in policy gradient method.

We have the blueprint for the learning agent to sample out the reliable control actions. But that is not only in the conventional working environments, where there are irregularities of the renewable energy supply. So we have to find a way to prove the controller is working for damping down the oscillations in time, in an environment which is prone to different uncertainties like that of renewable energy sources. With the special feature of the learning agent, uncertainties in the system can also be compensated accordingly. The unique design of the learning agent enables it to learn how to perform well in the power system setting, that is highly dynamic.

\begin{figure}[t]
\includegraphics[width=3.5in]{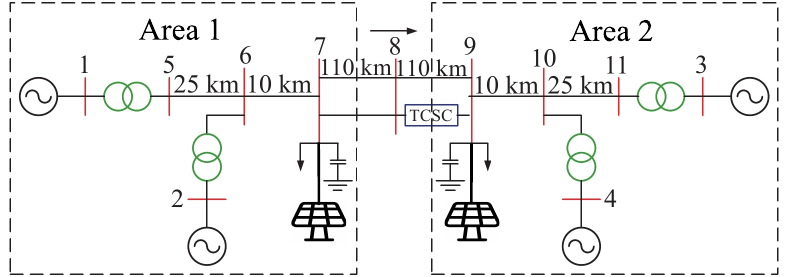}
\centering
\caption{Two-area (4-generator) system from \cite{kundur1994power} with additional renewable energy integration.}
\vspace{-4mm}
\label{fig:kundur}
\end{figure}
\begin{figure}[t]
\includegraphics[width=3in]{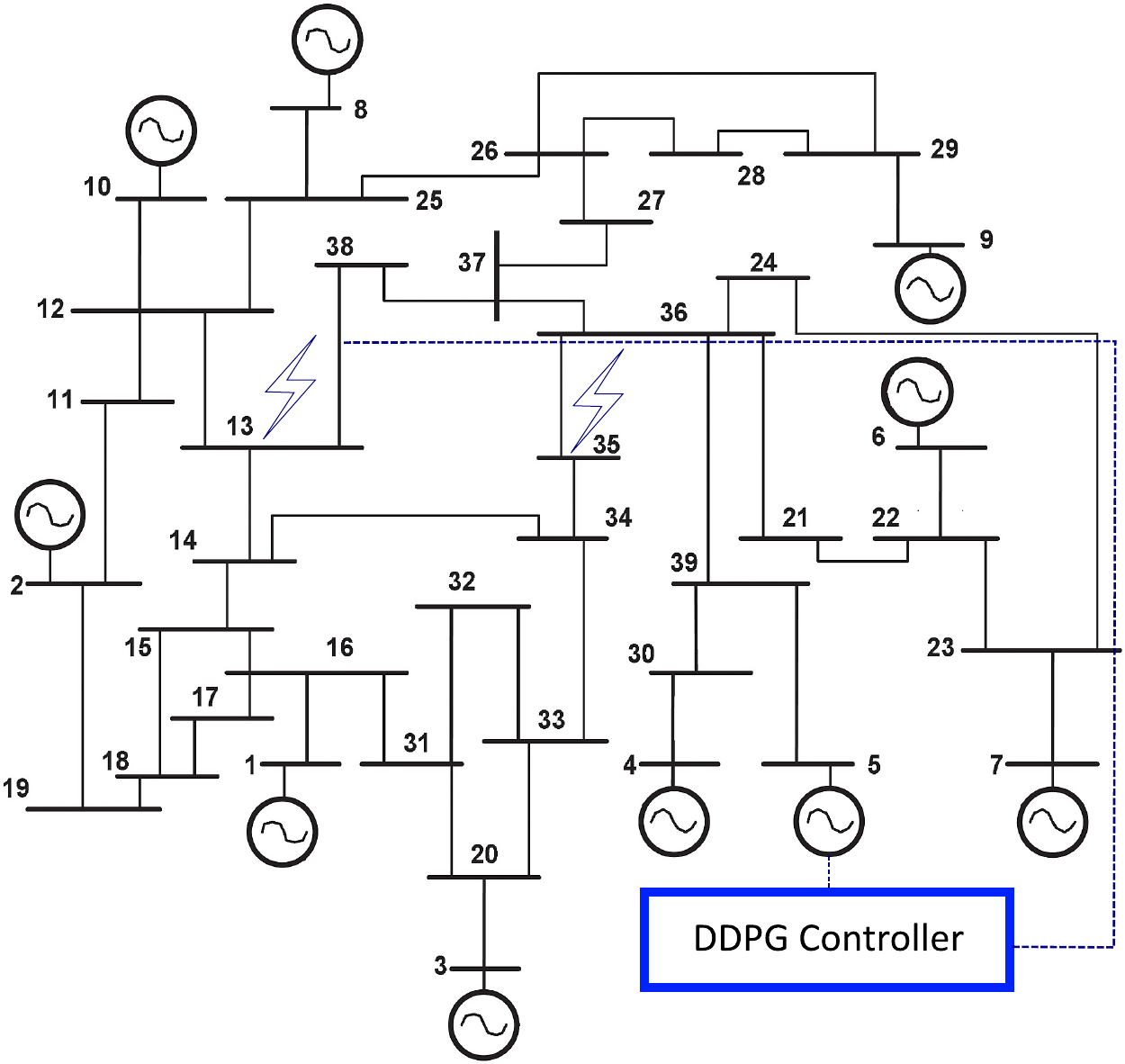}
\centering
\vspace{-4mm}
\caption{39-bus system with controller for the generator at bus 5. Drawing measurements from bus 13.}
\vspace{-4mm}
\label{fig:ddpg1}
\end{figure}
\begin{figure*}
\centering
    \centering
  \subfloat[\label{7a}]{%
       \includegraphics[width=0.2783\linewidth]{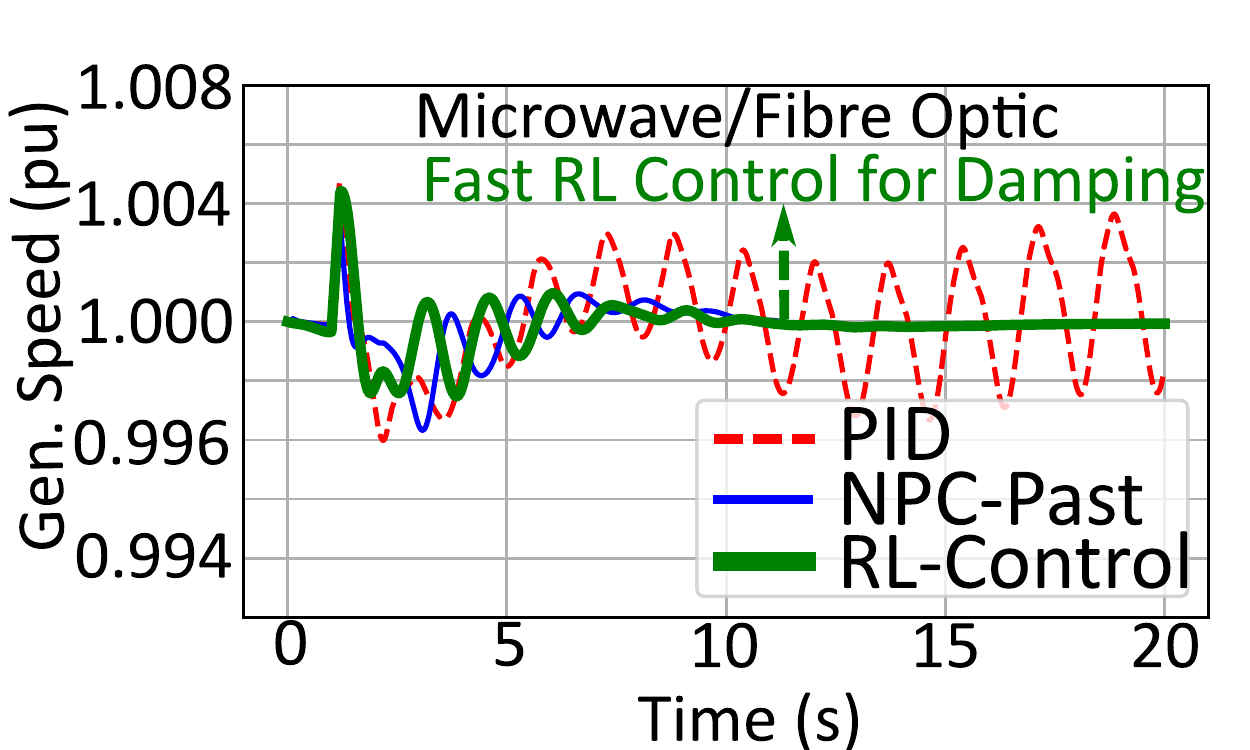}}
       \hfill
  \subfloat[\label{7b}]{%
        \includegraphics[width=0.23\linewidth]{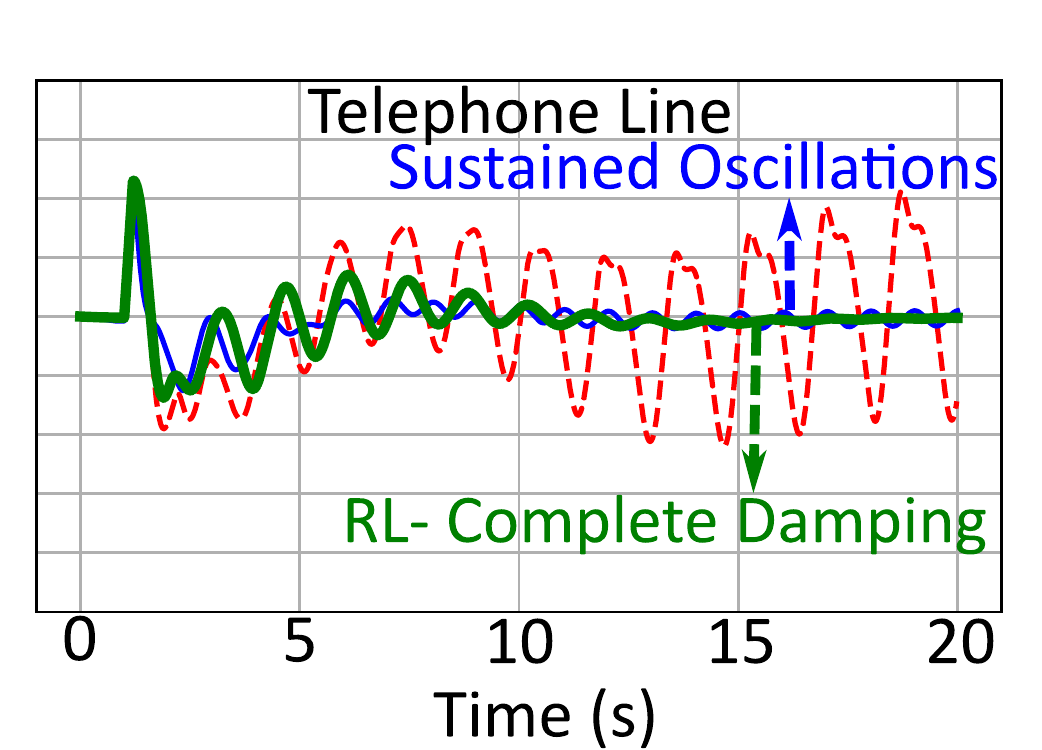}}
          \hfill
          \vspace{-4mm}
          \subfloat[\label{7c}]{%
        \includegraphics[width=0.231\linewidth]{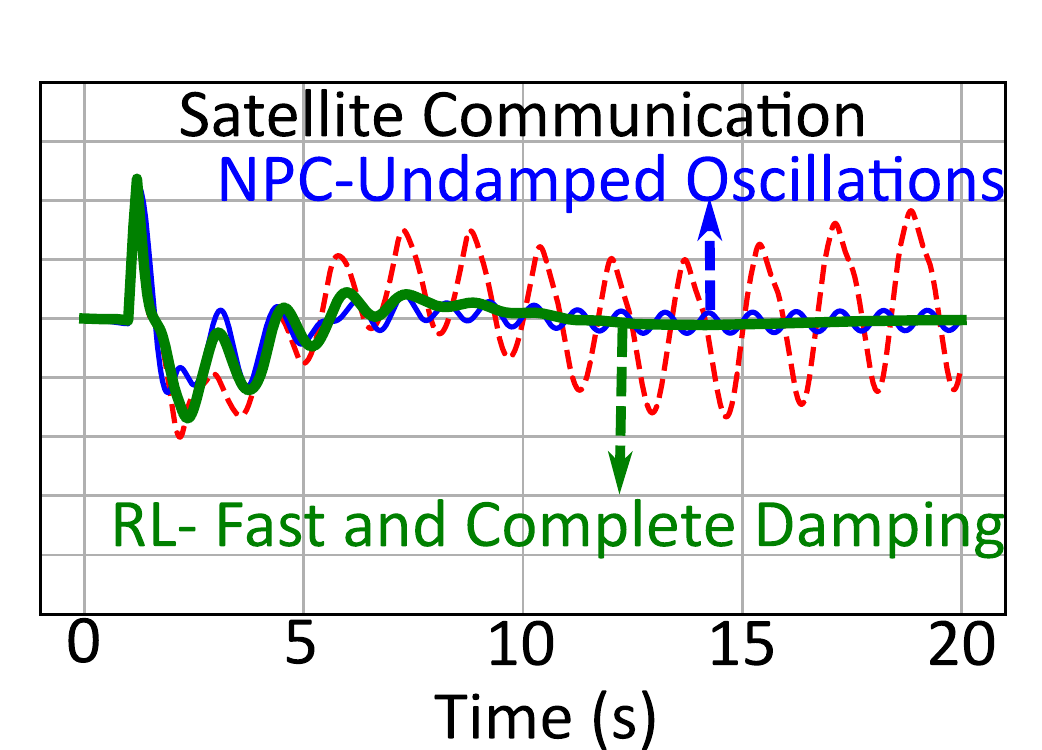}}
                 \hfill
                \subfloat[\label{7d}]{%
        \includegraphics[width=0.2365\linewidth]{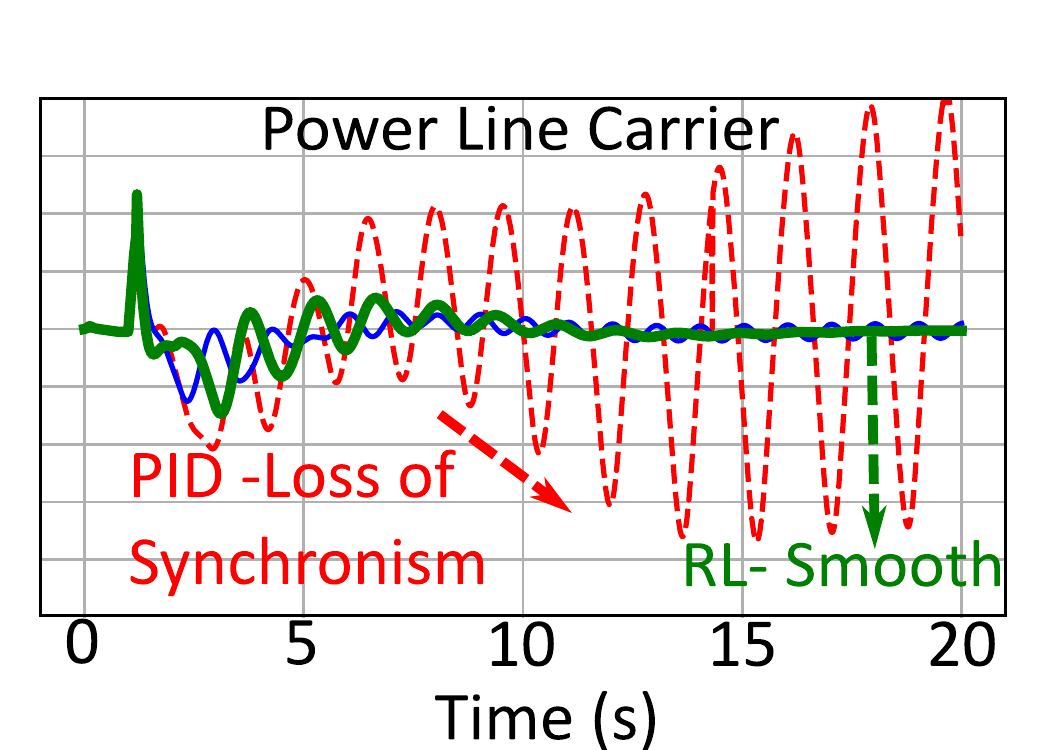}}
\vspace{3mm}
 \caption{Control method with reinforcement learning compared with past techniques with only the communication delay uncertainties corresponding to different communication channels.}
 \vspace{-4mm}
  \label{fig_rew_act_6} 

\end{figure*}

\begin{figure*}[ht]
\centering
    \centering
  \subfloat[\label{6a}]{%
       \includegraphics[width=0.273\linewidth]{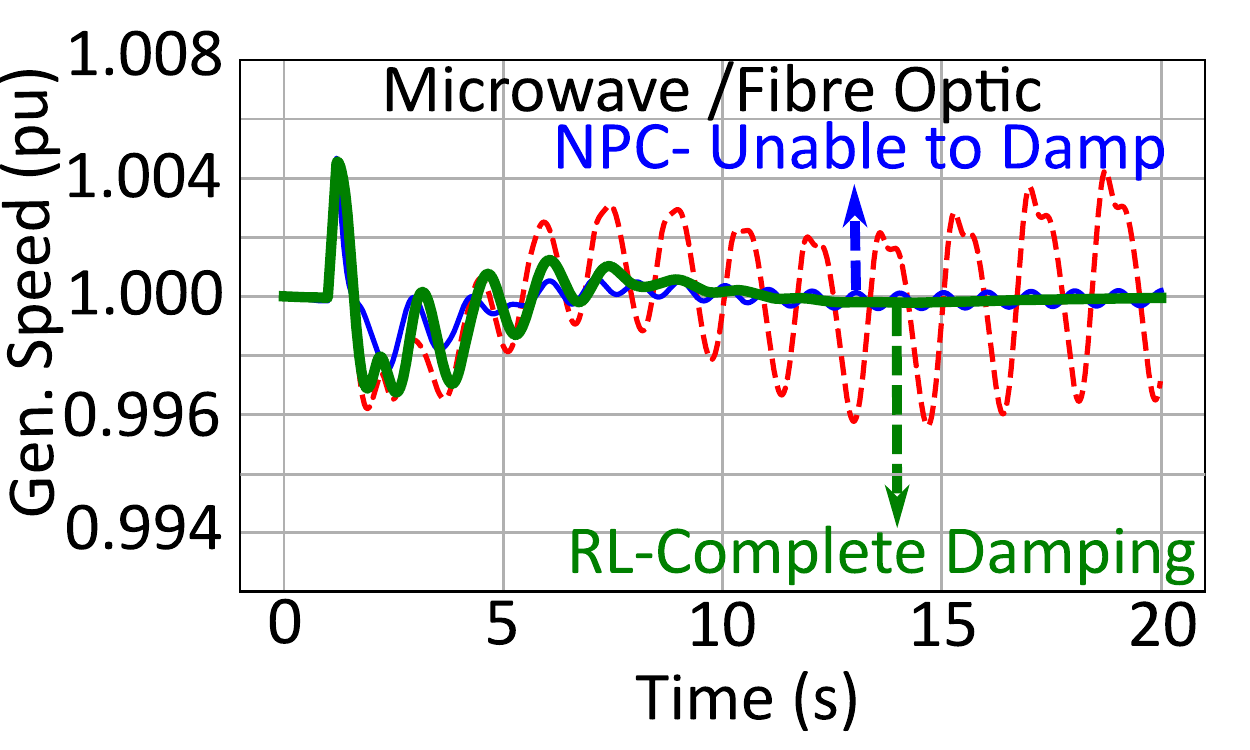}}
       \hfill
  \subfloat[\label{6b}]{%
        \includegraphics[width=0.232\linewidth]{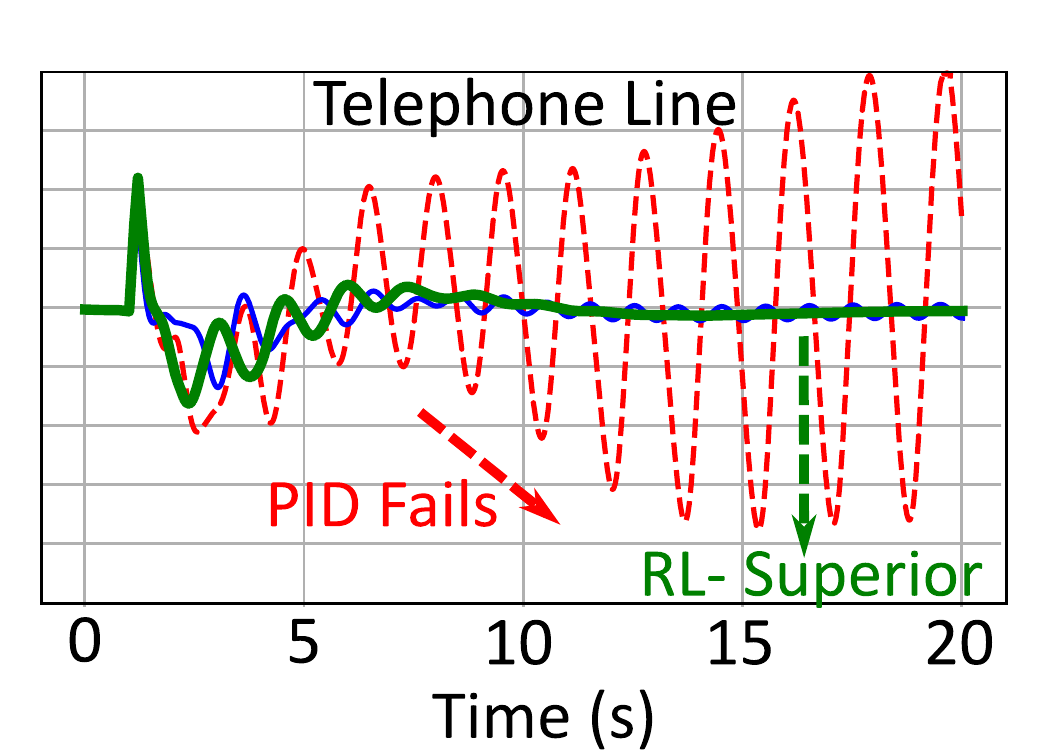}}
          \hfill
          \vspace{-4mm}
          \subfloat[\label{6c}]{%
        \includegraphics[width=0.2335\linewidth]{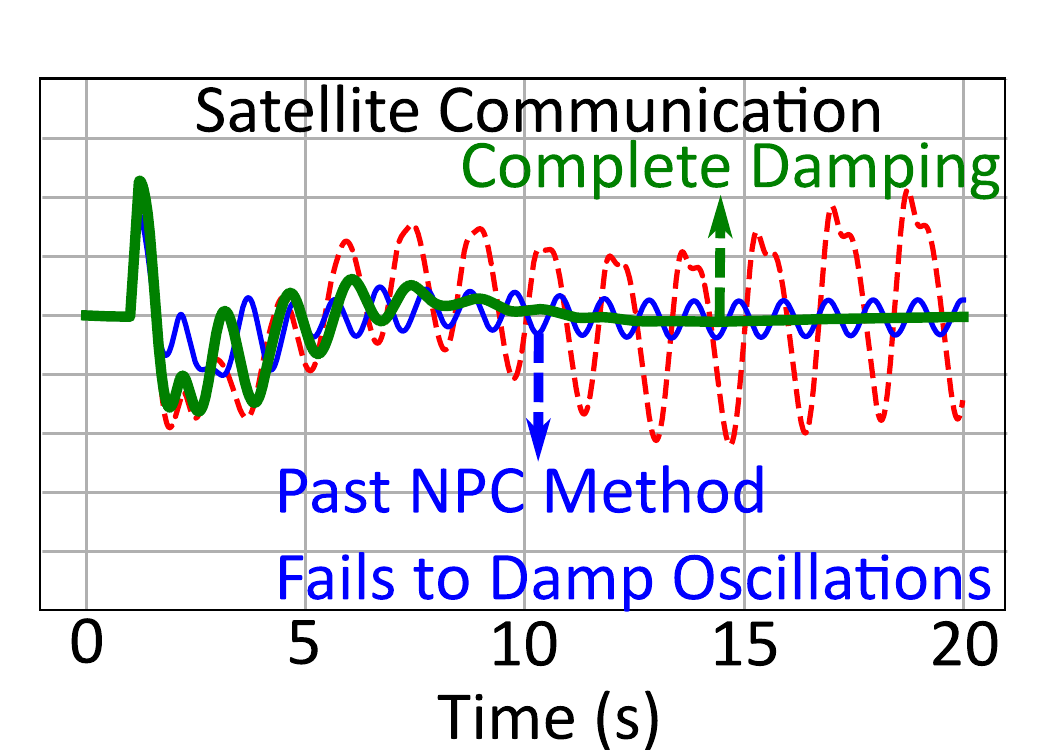}}
                 \hfill
                \subfloat[\label{6d}]{%
        \includegraphics[width=0.234\linewidth]{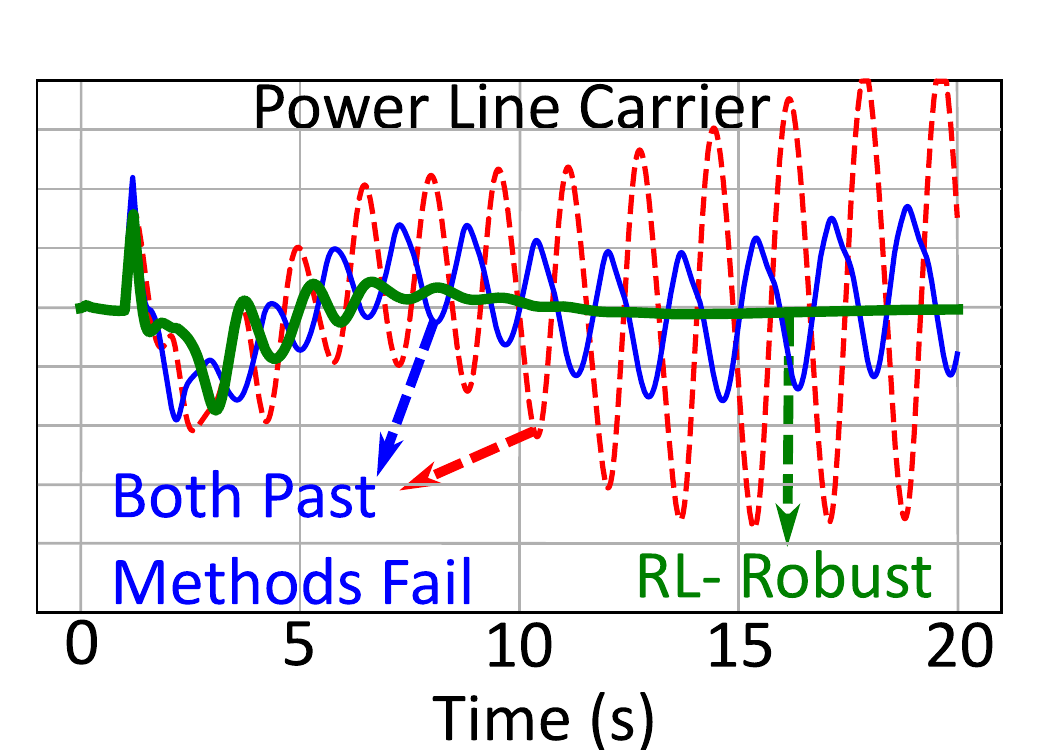}}
\vspace{3mm}
 \caption{Control method with reinforcement learning compared with past techniques when there are renewable energy uncertainties are introduced to the system along with the communication delay uncertainties corresponding to different communication channels.}
 \vspace{-6mm}
  \label{fig_rew_act_7} 
\end{figure*}

\section{Numerical Validation}

In this section, the proposal was applied to the 2-area and 4-generator Kundur system and the IEEE 39 Bus 10 generator system. A PID controller and a networked predictive controller (NPC) \cite{yao2014wide} are employed as benchmarks. 
\vspace{-4mm}
\subsection{Validation Setup}
For the small-signal studies, the $4$-generator model has been used extensively over the past to provide evidence to the methodologies adopted for transmission systems \cite{kundur1994power} and \cite{klein1992analytical}. 
The system under study consists of two symmetrical areas linked together by two $230$ kV lines of $220$ km long. It was specifically designed to study low-frequency electromechanical oscillations in large interconnected power systems. The detailed model is shown in Fig. \ref{fig:kundur}. We control all the $4$ generators through our proposed methodology because the limited computational capability allowed for a $4$-dimensional action space. 

Despite its small size,  behavior of typical systems in actual operation resemble this test case to a reasonable degree. Each area is equipped with two identical generating units rated $20$kV/$900$MVA \cite{kundur1994power} and \cite{klein1992analytical}. The synchronous machines have identical parameters, except for inertias which are $H = 6.5s$ in the first area and $H = 6.175s$ in the second area. Thermal plants having identical speed regulators are further assumed at all locations, besides to fast static exciters with a $200$ gain. The load is represented as constant impedances and split between the areas in such a way that area 1 is exporting $413$ MW to area $2$. Since the surge impedance loading of a single line is about $140$ MW, the system is somewhat stressed, even in steady-state. The reference load-flow with Area2 considered the slack machine is such that all generators are producing about $700$ MW each. Additionally, a solar power system of $100$ MW is added to both the areas. This creates an uncertain loads variation. 

The $10$-generator and 39 bus system from New England is also used to validate the proposed methodology, such a large system helps to establish the scalability of the mechanism. For the disturbance, we select the buses to apply faults that can give rise to oscillations. Since faults on any of the buses will have a similar effect on the controlled generators as long as at least one long-distance line exists in between the faulty bus and controlled generator. Keeping these points into consideration, we have a liberty to select bus $13$ and bus $35$, both of them have long-distance lines between them and controlled generator bus $5$. Similarly, the measurements are taken from the bus $13$ as the input to the reinforcement learning-based controller. The generator under reinforcement learning control is connected to bus $5$, as illustrated in Fig. \ref{fig:ddpg1}. The control mechanism is validated by controlling only one generator while the others are assumed to be controlled through standard PSSs. The reason for controlling only one generator is that the burden on computational capability becomes too high when action space becomes $10$ dimensional. The single generator control strategy is widely adopted for LFO damping control, as in \cite{4113518}.
\vspace{-4mm}
\subsection{Validation: Achieve Accurate Control Under Different Scenarios}
With a simulation environment in place, we interface the policy gradient-based reinforcement learning agent to the measurements from the state evaluator. The control action is supplied to the field voltage of the generators under consideration. The four generator model results are shown in Fig. \ref{fig_rew_act_7}, due to the space limitation.

Equations (\ref{rew1})-(\ref{eqn9}) show that the maximum attainable reward is $0$. The learning curve Fig. \ref{fig:ddpg10} shows that the model starts at a very low value and then upon learning on based on the discounted reward, the parameters of the neural networks are updated and the policy tends toward optimal value. After $5,000$ episodes, the average reward in Fig. \ref{fig:ddpg10} reaches a value close to $0$. Since the reward functions are dependent highly on the existence of the oscillations, we aim to show the performance of a well-learnt model when there are low-frequency oscillations, in case of different communication channel delays. Additionally, Fig. \ref{fig:ddpg10} validates the special reward design we proposed. The high penalty is enforced in the scenarios where the system loses synchronism since such a case will be responsible for a large outage of the system. We consider such scenarios as game-over for the model and there is no further evaluation performed so that the training time can be curtailed.

The model is exploring from episode $0$ to around episode $2,000$ as illustrated in Fig. \ref{fig:ddpg10}, because enough exploration is the primary requirement to learn an appropriate policy to effectively damp down the LFOs. The model achieves a high average reward after sufficient exploration. After $4,000$ episodes, the model converges to a very high reward. The effect of learning can also be observed from another perspective of implementing a control policy that maintains system synchronization for stable operation of the whole power system. Fig. \ref{fig:ddpg10} shows that after $3,400$ episodes that high fidelity control action enables the system to maintain stability.
\vspace{-4mm}
\subsection{Validation: Evaluate Performance Under Comm. Delays}
With a well-learnt model, our next step is to validate the performance of the model with communication delays. In Fig. \ref{7a} the performance of different algorithms with communication time delay in microwave link or fibre optic line is shown. The comparison is carried out among an optimized PID controller, a past method based on network predictive approach (NPC) , and out proposed reinforcement learning (RL)-based approach. It is safe to say that RL Control (reinforcement learning-based control) method has outperformed others. Although we prove the effectiveness of our model under one kind of communication channel, it is imperative to establish the effect of other communication channels available for SCADA measurements. Hence, Fig. \ref{7b} shows the comparison of the control schemes when the communication delays of telephone lines are simulated. Moreover, Fig. \ref{7c} and Fig. \ref{7d} show the performance of the proposed model for satellite link and PLC  respectively.


PLC has an estimated communication delay between $150$ ms to $300$ ms. Similar to the other channels, in this case, the proposed controller has the best performance. Moreover, the result in Fig. \ref{7d} shows that, the other methods have suffered worse performance, while the proposed method has successfully contained the oscillations in the case of satellite communication delay.

This shows that not only the model has achieved a high fidelity in controlling the generator but also, the timing delay randomness is successfully learnt. Whereas other methods like a tuned PID controller completely fails when there are significant delays in the communication channels, and the NPC method also suffers deterioration in performance. Hence, to accomplish a wide area-based controller, deep deterministic policy gradient provides the best solution under normal circumstances. 

\vspace{-4mm}

\subsection{Validation: Asses Efficacy Under PV Uncertainties}

In this section, we aim to provide validation of our proposed technique under the presence $50$\% of load shared by the solar power system in both the areas. We retrained the model, with time delay uncertainties and PV sources incorporated into the system. Fig. \ref{fig_rew_act_7} shows the frequency oscillations by different controllers under the influence of microwave or fibre optic link, telephone line satellite link, and power line carrier delays respectively, under the influence of renewable energy uncertainties. The results show that, however, the increase in uncertainty has created the control much difficult for the PID controller and the past methods, but it does not affect the RL Control much, because of the RL method can learn by exploring the environment. 

Specifically, for both telephone line and power line carrier communication channels, the tuned PID controller fails to provide a required control. Such a failure causes the speeds of the generators to deviate from the standard $1$ pu, and eventually causes the whole system to lose synchronism. For rest of the other two channels of microwave or fibre optic links and satellite links, the deviation of speeds is slower, but the end result will be the same, i.e., the loss of synchronism.

Moreover, the networked predictive control (NPC) shows a better performance than a simple PID controller. But with more uncertainties introduced into the system, the performance of NPC deteriorates as indicated in Fig. \ref{fig_rew_act_7}. The communication delays of microwave and fibre-optic that were well handled in the case without renewable energy integration, is not the case in the presence of renewable energy. The oscillations are not completely damped down in the case of microwave and fibre optic link as shown in Fig. \ref{6a}. 

The overall performance is not much different, under the delays due to other communication modes . That effect can be observed by comparing the results for telephone line, satellite communication and power line carrier modes. For both PID and NPC controllers, the low-frequency oscillations are not getting damped completely even after $20$ s for the telephone line and satellite communication. The case of power line carrier is even worse, where the oscillations keep growing in amplitude and that will result in the loss of synchronism for the whole system.
Hence, in the proposed method, the exploration-exploitation learning process with solar power systems can help the model to experience such scenarios as well and it can adjust the parameters to meet the requirement of damping down the oscillations. 

Whereas, the proposed method based on reinforcement learning has proven its reliability in successfully damping down the LFOs for all kinds of communication channels, even under the presence of uncertainties due to a deep penetration of solar power systems. For microwave and fibre optic link and telephone lines, the oscillations are damped down very quickly, whereas, the satellite and power line carriers have relatively slower damping. This proves that DDPG has the capability  of learning the uncertainties well, including those which are introduced due to communication delays, but also the ones introduced by renewable energy integration into the system.
\vspace{-4mm}
\subsection{Validation: Accomplish High-Speed Learning With Uncertainties}
\begin{figure}[t]
\includegraphics[width=3.5in]{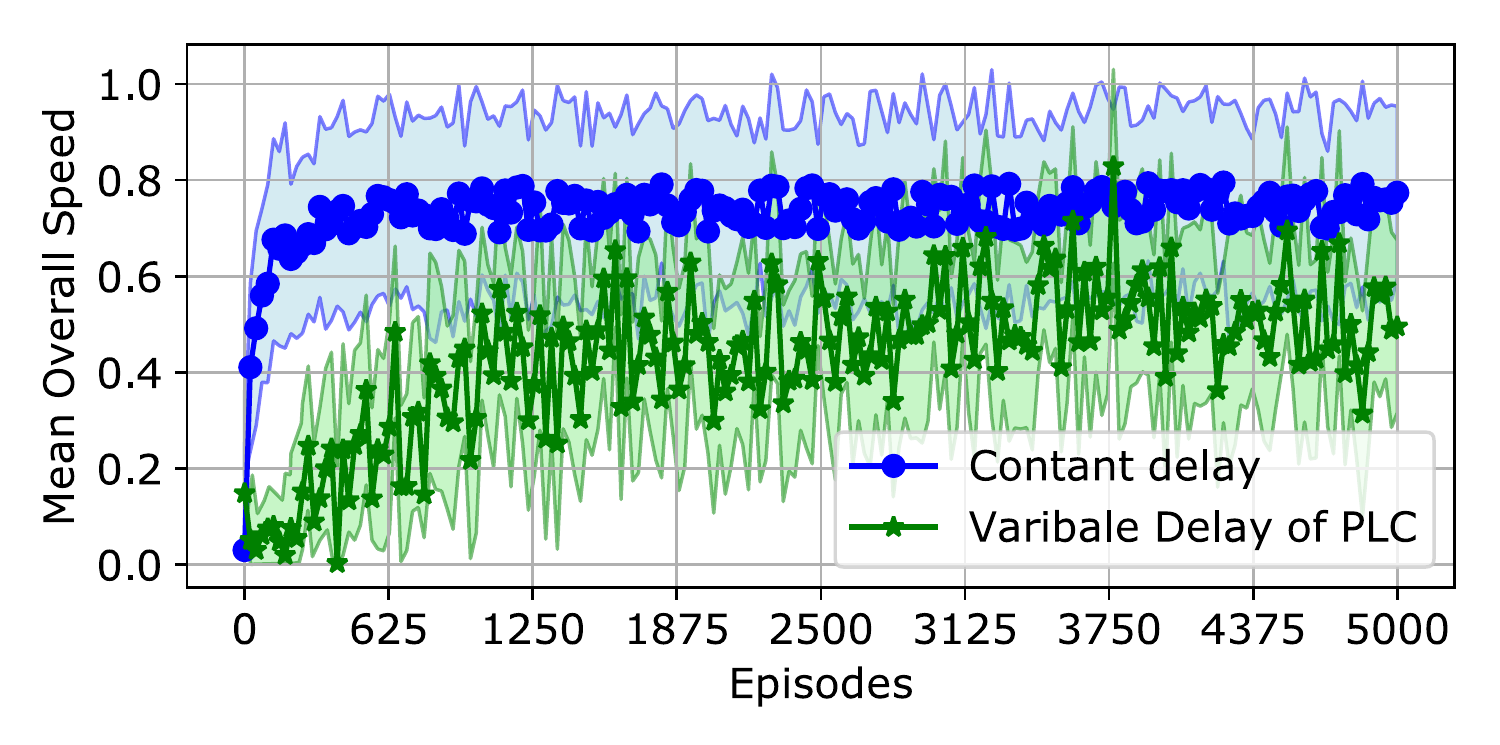}
\centering
\vspace{-7mm}
\caption{Mean overall speed $\bar{v}$ of the learning agent with constant and variable time delays by PLC.}
\vspace{-6mm}
\label{speed_of_learn}

\end{figure}

The accuracy of control methodology is not the only consideration while selecting the method, but researchers are also concerned with the speed and computational feasibility of the learning model. Therefore, we analyze the time required for learning when there is a constant time delay versus a variable uncertain time delay of the Gaussian mixture model. Fig. \ref{speed_of_learn} indicates, that more episodes are required to train the model when the communication delay has uncertainty. But it is a reasonable trade-off to spend that time to enable the model to be prepared for the uncertainties in the system so that we can achieve our goal of mitigating low-frequency oscillations without losing the observability of the system by leveraging wide area measurement systems.

\vspace{-4mm}
\section*{Conclusion}
The inter-area oscillations have been a cause of the blackouts and brownouts, but its control is a challenge that needs to be addressed with modern techniques. The wide area measurement systems provide a centralized control philosophy, however, it faces serious concerns of the uncertainties in the communication delays and the response of the equipment. We provide a holistic solution to such a problem by carefully modelling the control methodology and leverage the capability of learning the stochastic continuous control action through policy gradient method. Such a policy is learned by employing deep neural networks as the approximator. The method is validated numerically on numerous test cases under different scenarios of renewable energy penetration. The results prove the concept, scalability and robustness of the control system for low-frequency oscillations. Hence, a stable power system is ensured.
\vspace{-6mm}

\ifCLASSOPTIONcaptionsoff
  \newpage
\fi



%
\bibliographystyle{IEEEtran}
\bibliography{ref}

%




\end{document}